\documentclass[aps,prb,twocolumn,amsfonts,floatfix]{revtex4}

\usepackage{graphicx}% Include figure files
\usepackage{dcolumn}% Align table columns on decimal point
\usepackage{bm}% bold math
\usepackage{mathbbol}

\usepackage{amsmath,amssymb,amsthm}
\usepackage{times}
\usepackage{graphicx}
\usepackage{bm}
\usepackage[dvips]{color}

\begin{document}

\title{An introduction to the spectrum, symmetries, and dynamics of spin-1/2 Heisenberg chains}

\author{Kira Joel}
\author{Davida Kollmar}
\author{Lea F. Santos}
%\email{lsantos2@yu.edu}
\affiliation{Department of Physics, Yeshiva University, 245 Lexington Ave, New York, NY 10016, USA}

\begin{abstract}
Quantum spin chains are prototype quantum many-body systems. They are employed in the description of various complex physical phenomena. The goal of this paper is to provide an introduction to the subject by focusing on the time evolution of a Heisenberg spin-1/2 chain and interpreting the results based on the analysis of the eigenvalues, eigenstates, and symmetries of the system.  We make available online all computer codes used to obtain our data. 
\end{abstract}

\maketitle

\section{Introduction}

The fascination with magnets can be traced back as far as the antiquity in China, but it was only with the discovery of spins that we developed a better understanding of magnetism.~\cite{discoverSpin,Ohanian1986} Insulating solids with magnetic properties, in particular, can be viewed as lattices of atomic or ionic magnetic moments, each localized to one site.~\cite{MattisBook,NoltingRamakanthBook,ParkinsonBook} The total magnetic moment of the atom or ion depends, in general, on the spins of the electrons in incomplete shells and their orbital angular momenta. Here we refer to this total angular momentum simply as spin. The interactions between the spins may lead to collective behaviors with macroscopic effects, such as ferromagnetism, where the spins line up parallel to each other, and antiferromagnetism, where neighboring spins point in opposite directions. The source of such spontaneous magnetization is the so-called exchange interaction, introduced by Heisenberg and Dirac in the end of the 1920's.~\cite{MattisBook,NoltingRamakanthBook} This interaction has a quantum mechanical origin. It is the manifestation of the Coulomb repulsion between the electrons and the Pauli exclusion principle, being therefore strong and short range. A magnetic dipolar interaction is also present, but it is too small to explain magnetism at room temperature. 

The exchange interaction between particles of spin 1/2 and higher is usually described by the Heisenberg model. This is one of the most important models of magnetism and has been investigated for decades.~\cite{MattisBook,NoltingRamakanthBook,ParkinsonBook}  In 1931, Bethe found an exact analytical solution to the one-dimensional spin-1/2 Heisenberg model with coupling between nearest-neighbor sites.~\cite{Bethe1931,BetheIntro} This was a breakthrough in the studies of exactly solvable quantum many-body systems.~\cite{SutherlandBook} In 1983, Haldane suggested a remarkable difference between one-dimensional antiferromagnetic systems of integer and half-integer spins,~\cite{HaldanePLA1983,HaldanePRL1983} namely that the first should be gapped and the latter gapless.~\cite{Affleck1989} A necessary condition for a material to be insulating is the presence of an energy gap between the ground state and the first excited states. The prediction of the Haldane gap was confirmed numerically and experimentally, and a rigorous proof was soon provided for a similar model.~\cite{Affleck1987,Affleck1989}

The present work is restricted to the one-dimensional spin-1/2 Heisenberg model. Despite being a simplified theoretical model, it describes quantitatively well real materials found in nature or synthesized in laboratory, such as magnetic compounds. In some of these systems, unusual high transport of heat has been verified,~\cite{Sologubenko2000PRB,Hess2007,Hlubek2010} which has highly motivated the current interest in the subject. Such anomalous transport behavior has been associated with a macroscopic number of conserved quantities characterizing the model.~\cite{Zotos1997}

The spin-1/2 Heisenberg model finds applications in several other contexts. It is a key model in studies of quantum phase transition,~\cite{Chen2007} superconductivity,~\cite{MattisBook} localization in disordered systems,~\cite{Dukesz2009} as well as the dynamics~\cite{Santos2009JMP,Santos2011,Santos2012PRL} and thermalization~\cite{RigolARXIVBook} of correlated one-dimensional lattice systems. In quantum information, Heisenberg systems are used as models for quantum computers, each spin-1/2 representing a quantum bit (qubit), \cite{NielsenBook} in the analysis of entanglement, \cite{Amico2008} and in methods to transfer information in a controllable way.~\cite{Bose2007,Cappellaro2007} In the presence of impurities, disorder, or couplings beyond nearest-neighbors, the system becomes non-integrable and has been employed in the characterization of the crossover from integrability to quantum chaos.~\cite{Hsu1993,Avishai2002,Santos2004,Gubin2012}

There have also been attempts to simulate spin-1/2 Heisenberg chains with cold gases in optical lattices.~\cite{Simon2011}  Optical lattices are crystals of light. Laser beams propagating in opposite directions result in standing waves that confine ultracold atoms to small regions, the atoms playing the role of electrons in solid crystals.~\cite{Greiner2008} These systems are highly controllable, which allows for the simulation of condensed matter models not easily accessible with real solid state systems. Moreover, they are weakly coupled to the environment, which makes it possible to study their evolution for a long time. These factors combined make optical lattices essential tools to advance our understanding of quantum many-body systems far from equilibrium. The behavior of nonequilibrium systems is an outstanding challenge at the forefront of physics.

Motivated by the widely spread interest in spin systems and in the out-of-equilibrium properties of quantum many-body systems, we study here the factors that may limit the time evolution of the one-dimensional spin-1/2 Heisenberg model with couplings between nearest-neighbor sites only. We focus on the effects of an anisotropy parameter and on the symmetries of the system. We rely on the analysis of the eigenvalues and eigenstates of the system to anticipate its dynamics. Our predictions are then confirmed with actual numerical results for the time evolution. Since our studies require all eigenvalues and eigenstates, we use full exact diagonalization.

The paper is organized as follows. Section II provides a detailed description of the
Hamiltonian of the system. Section III analyzes the diagonal elements of the Hamiltonian, as well as its eigenvalues and eigenstates. Section IV investigates the time evolution of the system. Its symmetries and how they can constrain the system dynamics are discussed in Sec. V.

\section{Spin-1/2 chain}

The description of a system composed of a single spin-1/2 requires the use of the spin operators
$\hat{S}^{x,y,z} = \hat{\sigma}^{x,y,z}/2$, where 
\[ \hat{\sigma}^x \equiv \left(
\begin{array}{cc}
0 & 1  \\
1 & 0 
\end{array}
\right) ,
\hspace{0.5 cm}
 \hat{\sigma}^y \equiv \left(
\begin{array}{cc}
0 & -i  \\
i & 0 
\end{array}
\right) ,
\hspace{0.5 cm}
 \hat{\sigma}^z \equiv \left(
\begin{array}{cc}
1 & 0  \\
0 & -1 
\end{array}
\right)
\]
are the Pauli matrices and $\hbar$ has been set to 1. The quantum state of the spin is represented by a two-component vector, known as the spinor. This state is commonly written in terms of basis vectors corresponding to the eigenstates of $\hat{S}^z$, which are the spin pointing up in the $z$ direction $|\uparrow \rangle = \binom{1}{0}$ and the down-spin $|\downarrow \rangle = \binom{0}{1}$. The eigenvalue associated with $|\uparrow \rangle$ is +1/2 and that of  $|\downarrow \rangle$ is -1/2, which justifies referring to the first as the excitation. The operators $\hat{S}^{x}$ and $\hat{S}^{y}$ flip the up- or down-spin according to
\[
\hat{S}^{x} |\uparrow \rangle = \frac{1}{2} |\downarrow \rangle \hspace{0.7 cm} \hat{S}^{x} |\downarrow \rangle = \frac{1}{2} |\uparrow \rangle
\]
\[
\hat{S}^{y} |\uparrow \rangle = \frac{-i}{2} |\downarrow \rangle \hspace{0.7 cm} \hat{S}^{y} |\downarrow \rangle = \frac{i}{2} |\uparrow \rangle .
\]

Here, we study a one-dimensional system (chain) composed of $L$ coupled spins-1/2 described by the Heisenberg model, 
\begin{equation}
\hat{H} = \sum_{n=1}^{L-1} \left[ J \left(
\hat{S}_n^x \hat{S}_{n+1}^x + \hat{S}_n^y \hat{S}_{n+1}^y \right)  +
J_z  \hat{S}_n^z \hat{S}_{n+1}^z \right] .
\label{ham}
\end{equation}
The operators $\hat{S}^{x,y,z}_n$ act only on the spin placed on site $n$. The couplings are limited to nearest-neighbor spins; $J$ is the strength of the flip-flop term $\hat{S}_n^x \hat{S}_{n+1}^x + \hat{S}_n^y \hat{S}_{n+1}^y$, $J_z$ is the strength of the Ising interaction $\hat{S}_n^z \hat{S}_{n+1}^z$, and the ratio $\Delta=J_z/J$ is the anisotropy parameter. The model is isotropic when $\Delta=1$, in which case it is known as the XXX model, and it is anisotropic when $\Delta \neq 1$, usually referred to as the XXZ model (XYZ also exists when the coupling strengths in the three directions are different). A natural basis for the system is the set of $2^L$ states where the spin on each site is either pointing up or down, such as  $|\downarrow_{1} \uparrow_{2} \uparrow_{3} \dots \downarrow_{L} \rangle$. In quantum information, these states are known as quantum computational basis vectors. We refer to them as site-basis vectors.

We note that the one-dimensional spin-1/2 Heisenberg model can be mapped onto a system of hard-core bosons, that is bosons that cannot occupy the same site.~\cite{RigolARXIVBook} This system, in turn, is equivalent to the Bose-Hubbard model in the limit of strong repulsive interaction.~\cite{Cazalilla2011} The Bose-Hubbard model is used to describe interacting bosons on a lattice and the Hubbard model treats interacting fermions. The latter was introduced in 1963, \cite{Hubbard1963}  and since then these models have been extensively studied, especially for describing superconductivity and the transition from an insulator to a superfluid. \cite{Quintanilla2009,EsslerBook}

The flip-flop term in Hamiltonian~(\ref{ham}) interchanges the position of neighboring up and down spins according to 
\[
J(\hat{S}_n^x \hat{S}_{n+1}^x + \hat{S}_n^y \hat{S}_{n+1}^y)|\uparrow_n \downarrow_{n+1} \rangle = 
(J/2) |\downarrow_n \uparrow_{n+1} \rangle .
\]
It is also commonly written with raising $\hat{S}^{+} = \hat{S}^{x} + i \hat{S}^{y}$ and lowering $\hat{S}^{-} = \hat{S}^{x} - i \hat{S}^{y}$ spin operators,
\[
(J/2) (\hat{S}_n^+ \hat{S}_{n+1}^- + \hat{S}_{n+1}^+ \hat{S}_{n}^-)|\uparrow_n \downarrow_{n+1} \rangle = 
(J/2) |\downarrow_n \uparrow_{n+1} \rangle .
\]
The flip-flop term therefore couples site-basis vectors that differ only by the orientation of the spins in two adjacent sites. In this basis, it constitutes the off-diagonal elements of the Hamiltonian matrix. This term plays a key role in the evolution of the system by moving the excitations through the chain. 

In the case of open boundary conditions (open chain), as in Eq.~(\ref{ham}), where the sum goes from site $n=1$ to site $L-1$, an excitation on site 1 can move only to site 2 and from site $L$ to $L-1$. The scenario of a ring (closed chain), where an excitation on site $L$ can also move to site 1 corresponds to closed boundary conditions and will be discussed briefly in this paper.

The Ising interaction contributes to the diagonal part of the Hamiltonian matrix written in the site-basis. It causes a pair of adjacent parallel spins to have different energy from a pair of anti-parallel spins, because
\begin{equation}
J_z \hat{S}_n^z \hat{S}_{n+1}^z |\uparrow_n \uparrow_{n+1}\rangle =+(J_z/4) |\uparrow_n \uparrow_{n+1}\rangle,
\label{upup}
\end{equation}
while
\begin{equation}
J_z \hat{S}_n^z \hat{S}_{n+1}^z |\uparrow_n \downarrow_{n+1}\rangle =-(J_z/4) |\uparrow_n \downarrow_{n+1}\rangle.
\label{updown}
\end{equation}

A Hamiltonian containing only the Ising interaction, $\hat{H}_{zz}=J_z \hat{S}_n^z \hat{S}_{n+1}^z $, constitutes the Ising model and was employed in the first attempts to describe the phase transition from paramagnetism to ferromagnetism.~\cite{Cipra1987} As one can infer from Eqs.~(\ref{upup}) and (\ref{updown}), the ground state of this model depends on the sign of the interaction strength; it is ferromagnetic, with all spins aligned in the same direction, when $J_z<0$,  and it shows an antiferromagnetic arrangement with antiparallel neighboring spins when $J_z>0$. 

The state in which all spins align in the same direction is also an eigenstate of the Heisenberg model, because the flip-flop term has no effect on it. When the Heisenberg model is ferromagnetic ($J_z<0$), this state is a ground state. For the antiferromagnetic Heisenberg model ($J_z>0$), on the other hand, the ground state is more complicated than the simple configuration of antiparallel spins (details about it may be found, for instance, in \cite{ParkinsonBook}).

Here, we are interested not only in the ground state, but in all eigenvalues and eigenstates of the finite antiferromagnetic XXZ model with open boundary conditions. They are computed numerically and used to compare the static and dynamic properties of the system in two scenarios, when $\Delta \lesssim 1$ and when $\Delta \gg 1$. We address the role of the anisotropy parameter, as well as border effects and symmetries. 

Our Hamiltonian commutes with the total spin in the $z$ direction, ${\cal \hat{S}}^z=\sum_{n=1}^L \hat{S}_n^z$, that is, $[\hat{H},{\cal \hat{S}}^z]=0$. This means that the system is invariant by a rotation around the $z$-axis, or equivalently, it conserves ${\cal \hat{S}}^z$. As a result, the Hamiltonian matrix of a system with $L$ sites is composed of $L+1$ independent blocks (or subspaces), each with a fixed number $N\in [0,L]$ of up-spins. Therefore, even though the total dimension of the Hilbert space is $2^L$, we can diagonalize a single subspace at a time, each of dimension $D=\binom{L}{N}$. When $L$ is even, the largest subspace has $N=L/2$. In this case, full exact diagonalization can be carried out for $L\leq 14 $ with the {\it Mathematica} computer codes we provide.~\cite{www} For larger systems, we recommend using a high-level computer programming language, such as {\it Fortran} or {\it C$^{++}$}. Full exact diagonalizations have been performed for matrices with up to $D \sim 3 \times 10^4$. \cite{Santos2010PRE} 

Other symmetries may also be present, but this discussion is left for Sec.~V. In the next two sections, we focus on the analysis of the spectrum of a particular ${\cal \hat{S}}^z$-subspace and the structure of its eigenstates written in the site-basis, as a way to predict the dynamics of the system, and then confirm our expectations by investigating the actual time evolution of different initial states.

\section{Spectrum}

Before diagonalizing the XXZ Hamiltonian matrix, which we write in the site-basis, let us first look at its diagonal elements. They correspond to the eigenvalues of the Ising part, $\hat{H}_{zz}$, of the Hamiltonian~(\ref{ham}) and are split into sets of degenerate energies. Here, we refer to separated sets of energies as energy bands. The bands are determined by the number of pairs of adjacent parallel spins in the basis vectors. For each ${\cal \hat{S}}^z$-subspace, the larger the number of pairs, the larger the energy of the basis vector. For example, in an open chain with $L=4$ and $N=2$, the highest energy, $J\Delta/4$, occurs for the states with two pairs of parallel spins, $|\uparrow \uparrow \downarrow \downarrow \rangle$ and $|\downarrow \downarrow \uparrow \uparrow \rangle$. The band that precedes this one in energy has the states with only one pair of parallel spins, $|\uparrow  \downarrow \downarrow \uparrow \rangle$ and  $| \downarrow \uparrow \uparrow \downarrow \rangle$, yielding an energy of $-J\Delta/4$, while the states of the band with the lowest energy, $-3J\Delta/4$, have no pairs of parallel spins, $|\uparrow  \downarrow \uparrow \downarrow  \rangle$ and $| \downarrow \uparrow \downarrow \uparrow  \rangle$. One sees that the energy difference between consecutive bands is $J\Delta/2$, since we move down in energy by breaking a pair, thus adding the factor $J\Delta/4$ one less time and subtracting it one more time. In an open chain, where there are $L-1$ coupling bonds, the general expression for the energy of each band is therefore 
\begin{equation}
E_{zz}^{\text{open}} = [2 p - (L-1)] J\Delta/4,
\end{equation} 
where $p$ is the total number of pairs of adjacent parallel spins. 

In a closed chain, on the other hand, the energy difference between successive bands is $J\Delta$. In this case, there are $L$ bonds and always an even number of antiparallel pairs, because there is no border to absorb any of them. We move down in energy by breaking necessarily two pairs of parallel spins, the factor $J\Delta/4$ thus being added two less times and subtracted two more times. The diagonal energies are then given by 
\begin{equation}
E_{zz}^{\text{closed}} = [2 p - L] J\Delta/4.
\end{equation} 
Clearly, the closed chain has fewer bands than the open one, as shown in Table~\ref{table:numberofbands}  (Note: $\lfloor x \rfloor $ is the integer part of $x$).
\begin{table}[h]
\caption{Number of energy bands formed with the diagonal elements of the XXZ Hamiltonian written in the site-basis.}
\begin{center}
\begin{tabular}{ccc}
N & open & closed \\ [2pt]
\hline \\ [-7pt]
$<\lfloor \frac{L}{2} \rfloor$ & 2N & N \\ [2pt]
\hline \\ [-7pt]
$\lfloor \frac{L}{2} \rfloor$ and $\lfloor \frac{L+1}{2} \rfloor$ & (L-1) & N \\ [2pt]
\hline \\ [-7pt]
$>\lfloor \frac{L+1}{2}  \rfloor$ & 2(L-N) & (L-N) \\ [2pt]
\hline
\end{tabular}
\end{center}
\label{table:numberofbands}
\end{table} 

From now on, our analysis focuses solely on open chains. Notice, however, that extending the studies to the case of closed systems is straightforward and the codes for it are provided.~\cite{www} 

\begin{figure}[htb]
\includegraphics[width=0.4 \textwidth]{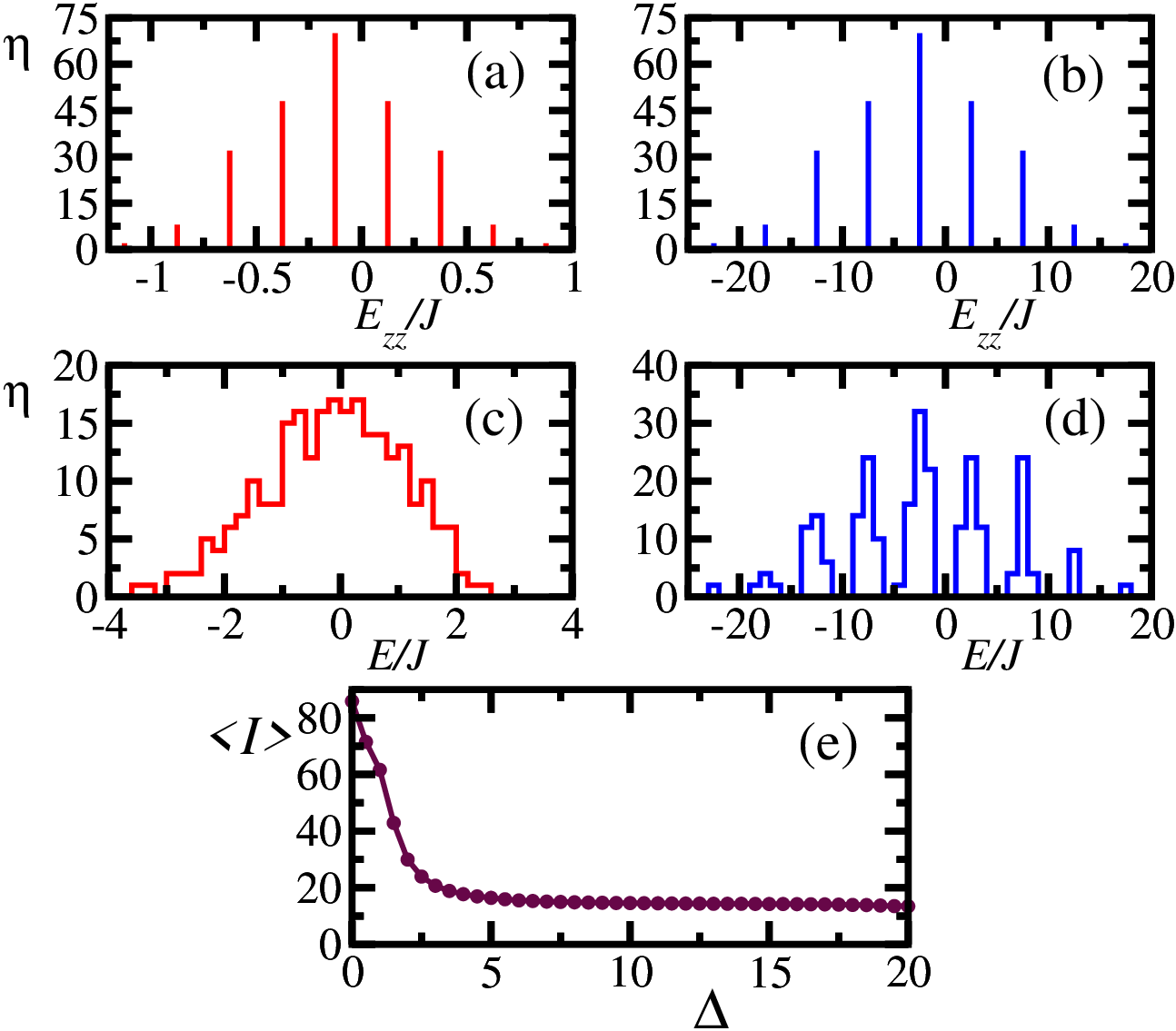}
\caption{(Color online) Histograms of the diagonal elements [panels (a) and (b)] and of the eigenvalues [panels (c) and (d)], and IPR averaged over all eigenstates vs $\Delta$ [panel (e)], for the antiferromagnetic XXZ Hamiltonian with open boundaries, $L=10$, and $N=5$. The site-basis is used. Panels (a) and (c): $\Delta=0.5$; panels (b) and (d): $\Delta=10$. Panel (c): bin width = 0.2 and panel (d): bin width = 1.0.}
\label{fig:histograms}
\end{figure}

For an open chain with an even number of sites and $N=L/2$, the diagonal elements of the XXZ Hamiltonian matrix form $L-1$ bands with energies ranging from $-(L-1)J\Delta/4$, when $p=0$, to $(L-3)J\Delta/4$, when $p=L-2$. The total number of states $\eta_B$ contained in an arbitrary band $B$, with $B\leq L/2$ being a positive integer, is equal to the number of states $\eta_{L-B}$ contained in the band $L-B$, that is the band structure is symmetric. 
The number of states in each band grows as we approach the middle of the spectrum. For example, 
\begin{eqnarray}
L=6: &&  \eta_1=\eta_{5}=2 \hspace{0.34 cm} \eta_2=\eta_{4}=4 \hspace{0.34 cm} \eta_3=6, \nonumber \\
L=8: &&   \eta_1=\eta_{7}=2 \hspace{0.34 cm} \eta_2=\eta_{6}=6 \hspace{0.34 cm} \eta_3=\eta_4=\eta_5=18, \nonumber \\ 
L=10: && \eta_1=\eta_{9}=2 \hspace{0.34 cm} \eta_2=\eta_{8}=8 \hspace{0.34 cm} \eta_3=\eta_7=32\nonumber \\
&& \eta_4=\eta_6=48 \hspace{0.34 cm} \eta_5=72. \nonumber 
\end{eqnarray}
The case for $L=10$ is illustrated  with histograms in Figs.~\ref{fig:histograms} (a) and (b) for two values of $\Delta$. The least populated bands are always the ones in the extremes containing only two states each, $\eta_1=\eta_{L-1}=2$. If $L \! \! \mod 4 \neq 0$, as in Fig.~\ref{fig:histograms}, $B=L/2$ is the most populated band, whereas if $L$ is divisible by 4, the three bands in the middle are the most populated with $\eta_{L/2-1}=\eta_{L/2}=\eta_{L/2+1}$. By studying the three examples above and larger system sizes, we arrived at the general equation,
\begin{equation}
\eta_B = \eta_{L-B}= 2 \prod_{k=1}^B \frac{N- \lfloor k/2\rfloor}{N\delta_{k,1}+\lfloor k/2\rfloor}.
\end{equation} 

Contrary to the diagonal elements, the eigenvalues of the total Hamiltonian (\ref{ham}) may or not form bands of energy depending on the interplay between the Ising interaction and the flip-flop term. Figures~\ref{fig:histograms} (c) and (d) show the histograms for the spectrum of the XXZ chain obtained with the same parameters considered in the top panels. In Fig.~\ref{fig:histograms} (c), where $\Delta \lesssim 1$, the band structure is lost. This happens because the energy difference between the basis vectors is $\lesssim J$, so the flip-flop term can couple intra- and also inter-band states, broadening significantly the range of energy values. In contrast, the band structure is preserved in Fig.~\ref{fig:histograms} (d). There, since $\Delta \gg 1$, states from different bands are too far off-resonance and the flip-flop term can effectively couple only states belonging to the same band. Each band then acquires a small width, which does not erase the energy gap between them.~\cite{substructure} 

The competition between the Ising and the flip-flop term of the Hamiltonian is reflected also in the structure of the eigenstates. As the Ising interaction increases, limiting the role of the flip-flop term, the eigenstates become less spread in the site-basis. This can be quantified, for example, with the so-called inverse participation ratio (IPR).~\cite{Izrailev1990,ZelevinskyRep1996} Consider an eigenstate 
\begin{equation}
|\psi^{(j)} \rangle =\sum_{k=1}^{D} a_{k}^{(j)} |\phi_k \rangle
\label{PsiBasis}
\end{equation} 
written in terms of arbitrary orthonormal basis vectors $|\phi_k\rangle$. IPR is defined as 
\begin{equation}
I^{(j)} \equiv \frac{1}{\sum_{k=1}^{D} |a_{k}^{(j)}|^4}.
\label{IPR}
\end{equation}
This quantity is proportional to the number of basis vectors which contribute to each eigenstate. It is small when the state is localized and large when the state is delocalized in the chosen basis. In our studies, $|\phi_k\rangle$ corresponds to the site-basis.

Figure~\ref{fig:histograms} (e) shows IPR averaged over all eigenstates, $\langle I \rangle$, for various values of the anisotropy parameter. The maximum delocalization occurs at $\Delta=0$. As the anisotropy increases,  $\langle I \rangle$ decays monotonically until the energy bands cease overlapping and $\langle I \rangle$ approaches a constant value.~\cite{Dukesz2009} In this latter scenario, the eigenstates become superpositions involving only intra-band basis vectors.

\section{Dynamics}

We now analyze the time evolution of different initial states, each corresponding to a specific site-basis vector, $|\Psi(0)\rangle = |\phi_k\rangle$. The source of the dynamics is the flip-flop term, which couples $|\phi_k\rangle$ with other states, transforming $|\Psi(t)\rangle$ into an evolving superposition of site-basis vectors. From the results for the eigenvalues and eigenstates described in the previous section, we expect the initial state to spread over several basis vectors when $\Delta \lesssim 1$, whereas the dynamics should be confined to states belonging to the same energy band as $|\Psi(0)\rangle$ when $\Delta \gg 1$. 

To confirm the above predictions, we study two quantities, the magnetization of each site,
\begin{equation}
M_n(t) = \langle \Psi(t) | \hat{S}^z_n| \Psi(t) \rangle ,
\end{equation}
and the probability $P_l(t)$ for finding a basis vector $|\phi_l\rangle $ at instants of times $t$. Since the Hamiltonian matrix (\ref{ham}) in the site-basis is real symmetric, one can find a set of orthonormal real eigenstates. This is indeed what we obtain from our numerical diagonalization. Furthermore, since the initial state is a single basis vector, we have $|\phi_k\rangle = \sum_{j=1}^D a_k^{(j)} |\psi^{(j)}\rangle$, where the coefficients $a$'s are real. This leads to
\begin{eqnarray}
|\Psi(t)\rangle &=& \sum_{j=1}^D a_k^{(j)} |\psi^{(j)}\rangle e^{-i E_j t} \nonumber \\
&=&  \sum_{l=1}^D \left( \sum_{j=1}^D a_k^{(j)} a_l^{(j)} e^{-i E_j t} \right) |\phi_l\rangle ,\nonumber
\end{eqnarray}
and the probability is then given by
\begin{equation}
P_l(t) = \left| \sum_{j=1}^D a_k^{(j)} a_l^{(j)} e^{-i E_j t} \right|^2.
\end{equation}

\begin{figure}[htb]
\includegraphics[width=0.4 \textwidth]{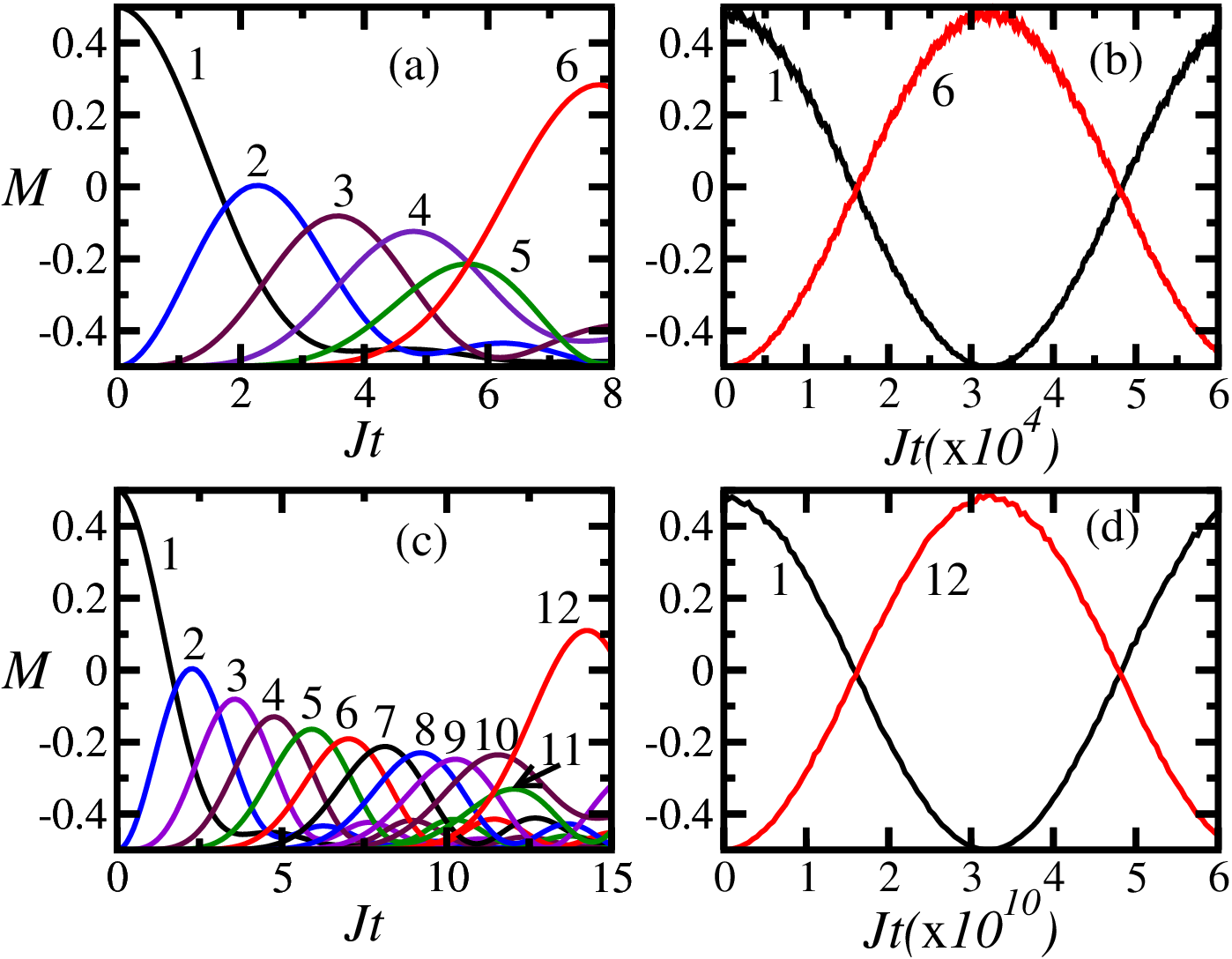}
\caption{(Color online) Magnetization of each site versus time; initial state $| \uparrow \downarrow \downarrow  \ldots \downarrow \rangle$. The sites are indicated with numbers.  Panels (a), (c): $\Delta=0.5$ and panels (b), (d): $\Delta=10$. Top panels: $L=6$ and bottom panels: $L=12$. All panels are for open chains.}
\label{fig:dynamics}
\end{figure}

The panels in Fig.~\ref{fig:dynamics} show the magnetization of each site for an open chain that has a single excitation initially placed on site 1, $|\Psi(0) \rangle=| \uparrow \downarrow \downarrow \ldots \downarrow \rangle $. In panels (a) and (c), where $\Delta =0.5$, the up-spin leaves the edge and gradually spreads through the chain by hopping successively from one site to the next in intervals of time $\sim J^{-1}$. The probability of finding it on a single site decreases from site 2 to $L-1$, but it finally reaches the other edge with high probability. The preference for the edges is a border effect that occurs only in open chains, because the border states $|\uparrow \downarrow \downarrow \ldots \downarrow \rangle $ and $|\downarrow \downarrow \ldots \downarrow \uparrow \rangle$ are in resonance. This effect decreases with system size, as seen by comparing panel (a), where $L=6$, with panel (c), where $L=12$. Since the energy difference between the border states and the states with the up-spin on sites $2\leq n\leq L-1$ is only $J\Delta/2=0.25J$, the latter can still take part in the dynamics. For comparison, we show in panels (b) and (d) the case where this energy difference is large, $J\Delta/2=5J$. There, only the border states are effectively coupled. The intermediate states, being so different in energy, have negligible participation in the evolution of the initial state. In perturbation theory, they are referred to as ``virtual states''. \cite{ShankarBook} The order of perturbation theory in which the border states are coupled is determined by the number of virtual states separating them, and their effective coupling strength, $ J_{\text{eff}}$, is inversely proportional to the order of perturbation theory and to the energy difference between coupled and intermediate states.  Therefore, the time $t\sim J_{\text{eff}}^{-1}$ for the excitation to move from $n=1$ to $n=L$ is long and increases with system size, as seen by the time scales in panels (b) and (d), where two different chain sizes are considered. 

\begin{figure}[htb]
\includegraphics[width=0.35 \textwidth]{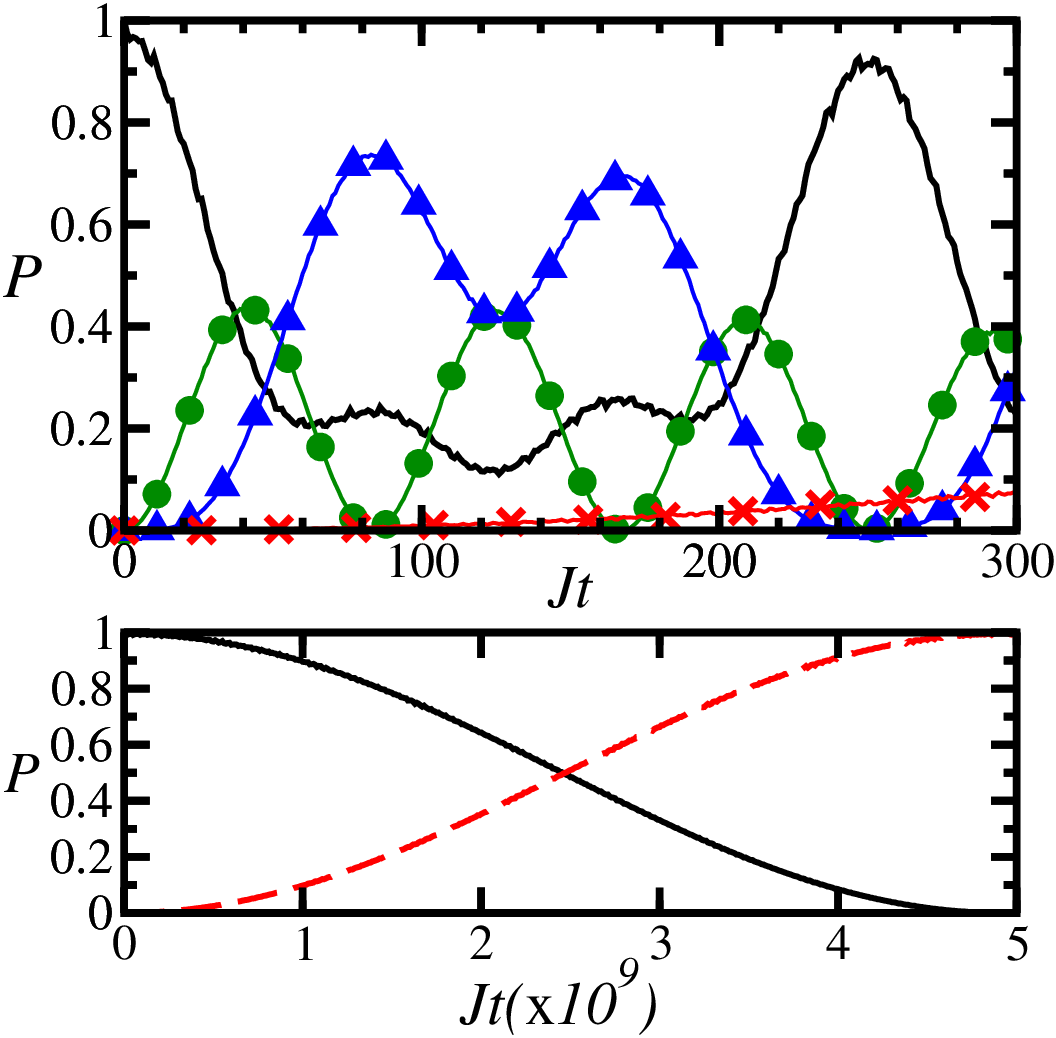}
\caption{(Color online) Probability in time to find a specific basis vector; open chain, $L=6$, and $\Delta=10$. Top panel: Initial state $| \downarrow \uparrow \uparrow \downarrow \downarrow \downarrow  \rangle $ (solid line) couples effectively with
 $| \downarrow \downarrow \uparrow \uparrow  \downarrow \downarrow  \rangle $ (circles) ,
 $| \downarrow \downarrow \downarrow \uparrow \uparrow   \downarrow  \rangle $ (triangles), and
 $| \uparrow \downarrow \downarrow \downarrow  \downarrow \uparrow \rangle $ (cross).
Bottom panel: initial state $| \uparrow \uparrow \uparrow \downarrow \downarrow \downarrow \rangle $ (solid line)
couples effectively only with $|\downarrow \downarrow \downarrow \uparrow \uparrow \uparrow \rangle  $ (dashed line).}
\label{fig:probability}
\end{figure}

The different time scales associated with the order of perturbation theory in which states are effectively coupled are well illustrated in the top panel of Fig.~\ref{fig:probability}. There we have 6 sites and 2 excitations initially placed away from the borders, but next to each other. Since the anisotropy considered is large, $\Delta=10$, the up-spins will tend to move through the chain as a bound pair. The figure shows the probability in time for each basis vector. The initial state $|\Psi(0) \rangle=| \downarrow \uparrow \uparrow \downarrow  \downarrow \downarrow \rangle$ (whose probability is indicated with a solid line) is in resonance with the basis vectors $| \downarrow \downarrow \uparrow \uparrow  \downarrow \downarrow  \rangle $ (circles) and $| \downarrow \downarrow \downarrow \uparrow \uparrow   \downarrow  \rangle $ (triangles), both of which have up-spins next to each other. It is also in resonance with state $| \uparrow \downarrow \downarrow \downarrow  \downarrow \uparrow \rangle $ (cross), which compensates for the absence of a bound pair of up-spins by placing each excitation on a border site. This resonance is therefore a border effect. The four states belong to the same energy band, which is well separated from the other bands. Thus, only these four states should be able to mix, as confirmed by the figure.

State $|\downarrow \uparrow \uparrow \downarrow  \downarrow \downarrow \rangle$ couples with state
$| \downarrow \downarrow \uparrow \uparrow  \downarrow \downarrow  \rangle $ in second order of perturbation theory via an intermediate transition where the pair splits into the virtual state $|\downarrow \uparrow \downarrow  \uparrow  \downarrow \downarrow \rangle$ and then recombines again. Since the energy difference $E_{|\downarrow \uparrow \uparrow \downarrow  \downarrow \downarrow \rangle} - E_{|\downarrow \uparrow \downarrow  \uparrow  \downarrow \downarrow \rangle} =J\Delta $, the effective coupling strength between $|\downarrow \uparrow \uparrow \downarrow  \downarrow \downarrow \rangle$  and $| \downarrow \downarrow \uparrow \uparrow  \downarrow \downarrow  \rangle $ obtained by perturbation theory  is $J_{\text{eff}} \sim J/\Delta$.~\cite{ShankarBook,MerzbacherBook}
The same coupling strength is found between states $| \downarrow \downarrow \uparrow \uparrow  \downarrow \downarrow  \rangle $ and $| \downarrow \downarrow \downarrow \uparrow \uparrow   \downarrow  \rangle $.
These states then hybridize at $t \sim \Delta J^{-1}$.  In contrast, the effective coupling between $|\downarrow \uparrow \uparrow \downarrow  \downarrow \downarrow \rangle$ and $| \uparrow \downarrow \downarrow \downarrow  \downarrow \uparrow \rangle $ occurs in fourth order of perturbation theory. It then takes much longer for the latter state to take part in the dynamics. As seen in the top panel, the probability to find this state is negligible for a long time. Moreover, similar to the discussion in Fig.~\ref{fig:dynamics}, this time further increases with system size and the maximum probability to ever find this state further decreases.

The combination of large anisotropy and border effects can slow down the evolution of the system significantly, to the point that an initial state may look stationary.~\cite{Santos2005exc,Santos2009JMP,Haque2010} This is the case shown in the bottom panel of Fig.~\ref{fig:probability}, where $\Delta =10$ and the initial state, $|\Psi(0) \rangle=| \uparrow  \uparrow \uparrow \downarrow \downarrow \downarrow \rangle$, has a bound triple of up-spins with one excitation on the border. The panel shows the probability in time for each basis vector. The initial state can hybridize only with  $| \downarrow \downarrow \downarrow \uparrow  \uparrow \uparrow \rangle$, because all the other states are very far in energy. The communication between the two states occurs in a very high order of perturbation theory, all three excitations taking a long time to cross the chain from the left to the right side. This makes the initial state look frozen for a long time. Notice that in a closed chain, where the borders are absent, the dynamics would be faster. In this case, the bound triple of excitations $\uparrow  \uparrow \uparrow$ would not be restricted to the edges, but would move together through the whole chain.

Separated energy bands caused by large anisotropy and the presence of borders can then limit the dynamics of the system to a portion of the Hilbert space. Another restrictive factor is the symmetries of the system, as we discuss next.

\section{Symmetries}

Operators that commute with the Hamiltonian have two important properties: their eigenstates are also eigenstates of $\hat{H}$, \cite{GriffithsBook} and they represent physical quantities that are conserved.~\cite{HenleyBook} The latter property comes from the fact that the expectation value of a conserved quantity, $\langle \hat{O} \rangle$, does not change in time, so
\[
\frac{d \langle \hat{O} \rangle}{dt} =\frac{i}{\hbar} \langle \hat{H}\hat{O} - \hat{O}\hat{H} \rangle = 0 \Rightarrow [\hat{H},\hat{O}]=0,
\]
The way to find such constants of motion is by looking for the symmetries of the system. According to Noether's theorem,~\cite{Noether} the invariance of the Hamiltonian under a symmetry operation must necessarily lead to a conserved quantity. For example, invariance of $\hat{H}$ under translation in
space leads to conservation of linear momentum, and invariance of $\hat{H}$ under translation
in time leads to conservation of total energy.

We have already encountered a conserved quantity of our system, the total spin in the $z$-direction. As we saw when studying the dynamics, the eigenvalue of ${\cal \hat{S}}^z$ for the initial state is conserved throughout the evolution. If the initial state has $N$ up-spins, then all the states that take part in its evolution must have the same number of up-spins. But our system shows additional symmetries, as we describe next.

Hamiltonian (\ref{ham}) is invariant under reflection, which leads to conservation of parity, that is, $\hat{H}$ commutes with the parity operator 
\[
\hat{\Pi} = \left\{
\begin{array}{ccc}
{\cal \hat{P}}_{1,L} {\cal \hat{P}}_{2,L-1} \ldots {\cal \hat{P}}_{\frac{L}{2},\frac{L+2}{2}} & \text{for} & L=\text{even}\\
{\cal \hat{P}}_{1,L}  {\cal \hat{P}}_{2,L-1} \ldots {\cal \hat{P}}_{\frac{L-1}{2},\frac{L+3}{2}} & \text{for} & L=\text{odd}
\end{array}
\right.
\]
where ${\cal \hat{P}}_{i,j}=(\hat{\sigma}^x_i \hat{\sigma}^x_j + \hat{\sigma}^y_i \hat{\sigma}^y_j + \hat{\sigma}^z_i \hat{\sigma}^z_j +\mathbb{1})/2$ is the permutation operator and $\mathbb{1}$ is the identity operator.  ${\cal \hat{P}}_{i,j}$ permutes the $i^{th}$ and $j^{th}$ vector spaces. For instance, $\hat{\Pi} |\uparrow \downarrow \downarrow \uparrow \downarrow\rangle = |\downarrow \uparrow \downarrow \downarrow \uparrow \rangle $. 

Invariance under reflection may be better understood by imagining a mirror at one edge of the chain. If parity is conserved, the probability of each basis vector in the eigenstate is equal to that of its reflection. For example, suppose we have $L=4$ and one excitation. The eigenstates of $\hat{H}$, which are also eigenstates of $\hat{\Pi}$, are
given by 
\[
|\psi^{(j)} \rangle = a_{1}^{(j)}|\uparrow \downarrow \downarrow \downarrow \rangle +
a_{2}^{(j)} |\downarrow \uparrow \downarrow \downarrow \rangle +
a_{3}^{(j)} |\downarrow \downarrow \uparrow \downarrow \rangle +
a_{4}^{(j)} |\downarrow \downarrow \downarrow \uparrow \rangle 
\] 
and the probability amplitudes are either $a_{1}^{(j)}=a_{4}^{(j)}$ and $a_{2}^{(j)}=a_{3}^{(j)} $ for even 
parity, $\Pi=+1$, or $a_{1}^{(j)}=- a_{4}^{(j)}$ and $a_{2}^{(j)}=- a_{3}^{(j)} $ for odd 
parity, $\Pi=-1$. [Notice that the hat in $\hat{\Pi}$ indicates the operator and its absence indicates the eigenvalue, $\hat{\Pi} |\psi^{(j)} \rangle = \Pi |\psi^{(j)} \rangle$.]

If $L$ is even and $N=L/2$, our Hamiltonian is also invariant under a global $\pi$ rotation around the $x$ axis. The operator that realizes this rotation is
\[
\hat{R}^x_{\pi} = \hat{\sigma}^x_1  \hat{\sigma}^x_2 \ldots \hat{\sigma}^x_L
\] 
and one can easily verify that $[\hat{H},\hat{R}^x_{\pi} ]=0$.
As an example, suppose we have $L=4$ and $N=2$. The eigenstate 
\begin{eqnarray}
|\psi^{(j)} \rangle &=& a_{1}^{(j)}|\uparrow \uparrow \downarrow \downarrow  \rangle +
a_{2}^{(j)} |\uparrow \downarrow \uparrow  \downarrow \rangle +
a_{3}^{(j)} |\uparrow \downarrow \downarrow \uparrow  \rangle \nonumber \\
&+&
a_{4}^{(j)} |\downarrow \uparrow \uparrow \downarrow  \rangle +
a_{5}^{(j)} |\downarrow \uparrow  \downarrow  \uparrow \rangle +
a_{6}^{(j)} |\downarrow \downarrow \uparrow \uparrow  \rangle , \nonumber
\end{eqnarray}
has either $a_{1}^{(j)}=a_{6}^{(j)}$, $a_{2}^{(j)}=a_{5}^{(j)}$, and $a_{3}^{(j)}=a_{4}^{(j)}$, in which case the eigenvalue of $\hat{R}^x_{\pi}$ is $R^x_{\pi}=+1$, or $a_{1}^{(j)}=- a_{6}^{(j)}$, $a_{2}^{(j)}=- a_{5}^{(j)}$, and $a_{3}^{(j)}=- a_{4}^{(j)}$, in which case $R^x_{\pi}=-1$.

There are two other symmetries, which we will not explore here. One is conservation of total spin, ${\cal \hat{S}}_T = \sum_n \vec{S}_n$, which occurs only in isotropic systems ($\Delta=1$), where $[\hat{H},{\cal \hat{S}}_T^2]=0$. The other is conservation of momentum, which happens in the closed chain due to its invariance by a translation in space. 

\begin{table}[htb]
\caption{The first column numerates the eigenstates $|\psi^{(j)} \rangle $ of an open XXZ chain with $\Delta=0.4$, $L=6$, and $N=3$. The second and third columns give, respectively, the eigenvalues of $\hat{\Pi}$ and $\hat{R}^x_{\pi}$ of these eigenstates. Fourth, fifth, sixth, and seventh columns: probability amplitudes $c_{A,B,C,D}^{(j)}$ of the initial states [Eq.~(\ref{initialABCD})] written as superpositions of the eigenstates of the Hamiltonian, 
$|\Psi_{A,B,C,D}(0) \rangle = \sum_j c_{A,B,C,D}^{(j)} |\psi^{(j)} \rangle $. The eigenvalues of $\hat{\Pi}$ and $\hat{R}^x_{\pi}$ for these initial states are shown in the second and third rows.}
\begin{center}
\begin{tabular}{ccc||cccc}
\multicolumn{4}{c} 
{}  &  &  &  \\ [2pt] \cline{4-7} \\[-7pt]
 & & & \hspace{0.1cm}  $|\Psi_A (0) \rangle $ \hspace{0.2cm}  & $|\Psi_B(0)\rangle $ \hspace{0.2cm} & $|\Psi_C(0) \rangle $ \hspace{0.2cm} & $|\Psi_D(0) \rangle $ \\ [2pt] \cline{4-7} \\[-7pt]
 & & &   $\Pi=+1$                            & $\Pi=-1$                            &         $\Pi=\O$               & $\Pi=-1$ \\ [2pt] 
 \cline{4-7} \\[-7pt]
 & & &      $R_{\pi}^x=\O$         &        $R_{\pi}^x=\O$               & $R_{\pi}^x=+1$      & $R_{\pi}^x=+1$ \\ [2pt] 
 \hline \\ [-7pt]
 &$\Pi$   &  $R_{\pi}^x$  & $c_A^{(j)}$    &   $c_B^{(j)}$  & $c_C^{(j)}$      & $c_D^{(j)}$   \\ [2pt] 
\hline \\ [-7pt]
$\psi^{(1)}$  & -- & -- &   0 & --0.16 & 0 & 0 \\ [2pt]
\hline \\ [-7pt]
$\psi^{(2)}$  & + & + &    --0.19 & 0 & --0.19 & 0 \\ [2pt]
\hline \\ [-7pt]
$\psi^{(3)}$  & -- & + &    0 & 0.33 & 0.33 & 0.46 \\ [2pt]
\hline \\ [-7pt]
$\psi^{(4)}$  & -- & -- &    0 & --0.07 & 0 & 0 \\ [2pt]
\hline \\ [-7pt]
$\psi^{(5)}$  & + & + &   0.19 & 0 & 0.19 & 0 \\ [2pt]
\hline \\ [-7pt]
$\psi^{(6)}$  & + & -- &    --0.48 & 0 & 0 & 0 \\ [2pt]
\hline \\ [-7pt]
$\psi^{(7)}$  & + & + &    --0.05 & 0 & --0.05 & 0 \\ [2pt]
\hline \\ [-7pt]
$\psi^{(8)}$  & -- & + &    0 & --0.28 & --0.28 & --0.40 \\ [2pt]
\hline \\ [-7pt]
$\psi^{(9)}$  & -- & -- &    0 & 0.33 & 0 & 0 \\ [2pt]
\hline \\ [-7pt]
$\psi^{(10)}$  & -- & -- &    0 & 0.42 & 0 & 0 \\ [2pt]
\hline \\ [-7pt]
$\psi^{(11)}$  & + & + &    --0.50 & 0 & --0.50 & 0 \\ [2pt]
\hline \\ [-7pt]
$\psi^{(12)}$  & + & -- &    --0.15 & 0 & 0 & 0 \\ [2pt]
\hline \\ [-7pt]
$\psi^{(13)}$  & + & + &    --0.34 & 0 & --0.34 & 0 \\ [2pt]
\hline \\ [-7pt]
$\psi^{(14)}$  & -- & -- &    0 & 0.11 & 0 & 0 \\ [2pt]
\hline \\ [-7pt]
$\psi^{(15)}$  & -- & -- &    0 & --0.30 & 0 & 0 \\ [2pt]
\hline \\ [-7pt]
$\psi^{(16)}$  & -- & + &    0 & --0.56 & --0.56 & --0.79 \\ [2pt]
\hline \\ [-7pt]
$\psi^{(17)}$  & + & -- &    0.50 & 0 & 0 & 0 \\ [2pt]
\hline \\ [-7pt]
$\psi^{(18)}$  & + & + &    0.04 & 0 & 0.04 & 0 \\ [2pt]
\hline \\ [-7pt]
$\psi^{(19)}$  & -- & -- &    0 & --0.28 & 0 & 0 \\ [2pt]
\hline \\ [-7pt]
$\psi^{(20)}$  & + & + &    --0.24 & 0 & --0.24 & 0 \\ [2pt]
\hline
\end{tabular}
\end{center}
\label{table:symmetry}
\end{table}
In this work, we focus on $\hat{\Pi}$ and $\hat{R}^x_{\pi}$ and analyze how they affect the dynamics of the system. For this, we consider four different initial states corresponding to superpositions of few basis-vectors,
\begin{eqnarray}
|\Psi_A(0) \rangle &=& (|\downarrow \uparrow \uparrow \uparrow \downarrow \downarrow \rangle + |\downarrow \downarrow \uparrow \uparrow \uparrow \downarrow \rangle )/\sqrt{2}, \nonumber \\
|\Psi_B(0) \rangle &=& (|\downarrow \uparrow \uparrow \uparrow \downarrow \downarrow \rangle - |\downarrow \downarrow \uparrow \uparrow \uparrow \downarrow \rangle )/\sqrt{2}, \nonumber \\
|\Psi_C(0) \rangle &=& (|\downarrow \uparrow \uparrow \uparrow \downarrow \downarrow \rangle + |\uparrow \downarrow \downarrow \downarrow \uparrow \uparrow \rangle )/\sqrt{2}, \nonumber \\
|\Psi_D(0) \rangle &=& (|\downarrow \uparrow \uparrow \uparrow \downarrow \downarrow \rangle - |\downarrow \downarrow \uparrow \uparrow \uparrow \downarrow \rangle +  \nonumber \\
&& \hspace{0.17 cm} |\uparrow \downarrow \downarrow \downarrow \uparrow \uparrow  \rangle  - |\uparrow \uparrow \downarrow \downarrow \downarrow \uparrow \rangle)/2.
\label{initialABCD}
\end{eqnarray}
The first two states are not eigenstates of $\hat{R}^x_{\pi}$, but parity is well defined, $|\Psi_A(0) \rangle$ has even parity and $|\Psi_B(0) \rangle$ has odd parity. State $|\Psi_C(0) \rangle$, on the other hand, is an eigenstate of the operator $\hat{R}^x_{\pi}$ with eigenvalue $+1$, but not of $\hat{\Pi}$. The last state, $|\Psi_D(0) \rangle$ is an eigenstate of both $\hat{\Pi}$ and $\hat{R}^x_{\pi}$, with eigenvalues $-1$ and $+1$, respectively.

In Table~\ref{table:symmetry}, we write each initial state as a linear superposition of the eigenstates of the system, $|\Psi_{A,B,C,D}(0) \rangle = \sum_j  c_{A,B,C,D}^{(j)} |\psi^{(j)} \rangle $, so that we can investigate which eigenstates can take part in the evolution. The first column numerates the eigenstates, $|\psi^{(j)} \rangle $. The second and third columns give the eigenvalues of $\hat{\Pi}$ and $\hat{R}^x_{\pi}$, respectively, for each eigenstate. The fourth, fifth, sixth, and seventh columns show the values of the probability amplitudes, $c_{A,B,C,D}^{(j)}$, of the eigenstates for the initial states $|\Psi_A (0) \rangle $, $|\Psi_B (0) \rangle $, $|\Psi_C (0) \rangle $, and $|\Psi_D (0) \rangle $, respectively. The eigenvalues of $\hat{\Pi}$ and $\hat{R}^x_{\pi}$ for each initial state are shown in the second and third rows.

From the table, we can see that only the eigenstates with the same symmetries as the initial state can contribute to the evolution of the latter. For $|\Psi_A (0) \rangle $ and $|\Psi_B (0) \rangle $, eigenstates with both values of $R_{\pi}^x$ are seen, but parity is strictly conserved. In the first case, the probability amplitudes of all odd eigenstates are zero and for the second state, $c_B^{(j)}=0$ for all eigenstates with $\Pi=+1$. For state $|\Psi_C (0) \rangle $, eigenstates with both parities are part of the superposition, but $c_C^{(j)}\neq 0$ only for eigenstates with $R_{\pi}^x=+1$. The last state, $|\Psi_D (0) \rangle $, has both symmetries, so only three of the 20 eigenstates have $c_D^{(j)}\neq 0$, those with $\Pi=-1$ and  $R_{\pi}^x=+1$ simultaneously. 

The three eigenstates constituting $|\Psi_D (0) \rangle $ are $|\psi^{(3)} \rangle, |\psi^{(8)} \rangle$, and $|\psi^{(16)} \rangle$. They are more localized in the site-basis than all other eigenstates, {\em i.e.}  they have fewer coefficients $a_k^{(j)} \neq 0$ in  Eq.~(\ref{PsiBasis}). This is because superpositions of some of the basis-vectors cannot satisfy both $\Pi=-1$ and  $R_{\pi}^x=+1$ at the same time. For example, the effect of the two operators $\hat{\Pi}$ and  $\hat{R}_{\pi}^x$ on $b_1 |\uparrow \uparrow \uparrow \downarrow \downarrow  \downarrow \rangle + b_2 | \downarrow \downarrow  \downarrow \uparrow \uparrow \uparrow \rangle $ is the same, so they cannot give different eigenvalues. If $\Pi=-1$, we necessarily have $b_2=-b_1 $, for which case $R_{\pi}^x$ is inevitably -1. These two basis vectors (and similarly for others) cannot, therefore, be part of the contributing eigenstates $|\psi^{(3)} \rangle, |\psi^{(8)} \rangle$, and $|\psi^{(16)}\rangle$; the coefficients $a_k^{(3),(8),(16)}$ associated with these basis vectors can only be zero.

In summary, the larger the number of conserved quantities, the smaller the invariant subspaces. The time evolution of an initial state with many constants of motion is therefore more constrained in the Hilbert space. 

\vspace{0.7cm}

\section{Discussion}

We make available online all computer codes used to obtain our data, as well as detailed explanations about them.~\cite{www}. Students and professors should have no difficulty to reproduce our results and to explore further questions. The studies described here can constitute an entire summer project, as in our case, can set the basis for a senior thesis, or give ideas for assignments in courses about Quantum Mechanics. By the time of completion of this work, the two first authors of this paper were undergraduate students who had not had a course on Quantum Mechanics, but had a solid background on Linear Algebra.

\centerline{\bf Acknowledgments}
%\vskip 0.1 cm
%\begin{acknowledgments}
This work was supported by the NSF under Grant DMR-1147430 and in part under Grant PHY11-25915 and PHY05-51164.
D.~K. thanks Stern College for Women of Yeshiva University for a summer fellowship. K.~J. and D.~K. thank the Kressel Research Scholarship for a one-year financial support.
This work is part of their theses for the S. Daniel Abraham Honors Program. L.~F.~S. thanks the KITP at UCSB for its hospitality.
%\end{acknowledgments}

\end{document}